\documentclass[lettersize,journal]{IEEEtran}
\usepackage{amsmath,amsfonts}
\usepackage{algorithmic}
\usepackage{array}
\usepackage[caption=false,font=normalsize,labelfont=sf,textfont=sf]{subfig}
\usepackage{textcomp}
\usepackage{stfloats}
\usepackage{url}
\usepackage{verbatim}
\usepackage{graphicx}
\usepackage{cite}
\usepackage{multirow}
\usepackage{float}
\usepackage{amsfonts,amssymb}
\usepackage{amsthm}
\usepackage{tabularx}
\usepackage{bbding}
\usepackage{color}
\usepackage{booktabs}
\usepackage{hyperref}
\usepackage{verbatim}
\usepackage[linesnumbered,lined,boxed,commentsnumbered,ruled]{algorithm2e}
\begin{document}

\title{Speech Separation with Pretrained Frontend to Minimize Domain Mismatch}
\author{Wupeng Wang,Zexu Pan,~\IEEEmembership{,Member,~IEEE}, Xinke Li,Shuai Wang,~\IEEEmembership{,Member,~IEEE},and Haizhou Li,~\IEEEmembership{,Fellow,~IEEE}

\thanks{
The research is supported by National Natural Science Foundation of China (Grant No. 62271432), Shenzhen Science and Technology Program ZDSYS20230626091302006, and Shenzhen Science and Technology Research Fund (Fundamental Research Key Project Grant No. JCYJ20220818103001002). (\textit{Corresponding author: Shuai Wang}).}
\thanks{Wupeng Wang is with the Department of Electrical and Computer Engineering, National University of Singapore, 119077 Singapore (e-mail: e0125301@u.nus.edu).}
\thanks{Zexu Pan is with the Alibaba Group, 189554 Singapore (e-mail: zexu.pan@alibaba-inc.com).}
\thanks{Xinke Li is with the Industrial Systems Engineering and Management, National University of Singapore, 119077 Singapore (e-mail: xinke.li@u.nus.edu).}
\thanks{Shuai Wang and Haizhou Li are with the Shenzhen Research Institute of Big Data, Chinese University of Hong Kong, Shenzhen, Guangdong, 518172 P.R.
China (e-mail: wangshuai@cuhk.edu.cn; haizhouli@cuhk.edu.cn).}

}

\markboth{Journal of \LaTeX\ Class Files,~Vol.~14, No.~8, August~2021}%
{Shell \MakeLowercase{\textit{et al.}}: A Sample Article Using IEEEtran.cls for IEEE Journals}


\maketitle
\begin{abstract}

Speech separation seeks to separate individual speech signals from a speech mixture.  Typically, most separation models are trained on synthetic data due to the unavailability of target reference in real-world cocktail party scenarios. As a result, there exists a domain gap between real and synthetic data when deploying speech separation models in real-world applications. In this paper, we propose a self-supervised domain-invariant pretrained (DIP) frontend that is exposed to mixture data without the need for target reference speech. The DIP frontend utilizes a Siamese network with two innovative pretext tasks, mixture predictive coding (MPC) and mixture invariant coding (MIC), to capture shared contextual cues between real and synthetic unlabeled mixtures. Subsequently, we freeze the DIP frontend as a feature extractor when training the downstream speech separation models on synthetic data. By pretraining the DIP frontend with the contextual cues, we expect that the speech separation skills learned from synthetic data can be effectively transferred to real data. To benefit from the DIP frontend, we introduce a novel separation pipeline to align the feature resolution of the separation models. We evaluate the speech separation quality on standard benchmarks and real-world datasets. The results confirm the superiority of our DIP frontend over existing speech separation models. This study underscores the potential of large-scale pretraining to enhance the quality and intelligibility of speech separation in real-world applications.
\end{abstract}


\section{Introduction}
\label{sec:introduction}
\IEEEPARstart{S}{peech} separation seeks to separate individual speech signals in a complex, multi-speaker acoustic environments, commonly known as the `cocktail party problem'~\cite{bronkhorst2000cocktail}. Speech separation has been an active research, not only because it is a human cognitive ability, but also it serves as a frontend in many practical speech processing tasks, such as speaker recognition~\cite{rao2019target,xu2021target, liu2022neural}, automatic speech recognition~\cite{yue2019end,wang2020voicefilter,gabrys2022voice}, and voice conversion~\cite{zhou2020multi,zhou2021seen}.

Traditional speech separation technique includes independent component analysis (ICA)~\cite{lee1998independent,araki2004underdetermined,choi2005blind}, non-negative matrix factorization (NMF)~\cite{wang2014discriminative,weninger2014discriminative,virtanen2007monaural} or the theory of computational auditory scene analysis (CASA)~\cite{lyon1983computational,wang2006computational,hu2007auditory}. With the advent of deep learning, supervised speech separation models, including both time-domain~\cite{luo2019conv,luo2020dual,subakan2021attention} and frequency-domain models~\cite{hershey2016deep,kolbaek2017multitalker,liu2019divide}, have achieved remarkable advancements. Despite impressive results, these models don't generalize well to real-world scenarios. This is primarily due to the fact that most speech separation models are trained on synthetic anechoic data~\cite{cosentino2020librimix}, that are very different from real-world speech data with unpredictable attenuation and reverberation~\cite{nagrani2017voxceleb,carletta2005ami,subakan2022real}. In this paper, such a problem is referred to as domain mismatch.

To mitigate this performance degradation due to domain mismatch, there were attempts to incorporate real-world speech mixtures during training. Unsupervised speech separation is one example. In~\cite{wisdom2020unsupervised}, the authors utilize the mixture waveform to generate the mixture of the mixture (MOM) and train the separation model with the mixture invariant training method. Tzinis et al.~\cite{tzinis2022remixit} expand this architecture by incorporating various remixing methods. Karamatli et al.~\cite{karamatli2022mixcycle} further promote the model through cyclic learning for continuous improvement. Semi-supervised speech separation is another example. Zhang et al.~\cite{zhang2021teacher} introduce a framework where the student model learns to predict pseudo-labels from the teacher model. In~\cite{han2023heterogeneous}, Han et al. extend this approach by introducing an ensemble structure combined with separation consistency training methods. Furthermore, Sivaraman et al.~\cite{sivaraman2022adapting} propose a combination of semi-supervised and unsupervised methods to improve the model performance in meeting scenarios. 

The above techniques incorporate real speech mixtures during model training. However, the quality improvement over conventional supervised methods~\cite{wisdom2020unsupervised} is limited, probably because they introduce another mismatch, i.e. target mismatch. The target labels in those techniques are either the mixture itself or fake labels rather than the clean reference speech. In a different attempt, Ben et al.~\cite{ben2006analysis} demonstrate that the domain-invariant knowledge is crucial for models to overcome domain mismatch. The effectiveness of domain-invariant knowledge has also been validated in various tasks~\cite{thota2021contrastive, zhu2022multi,aharoni2020unsupervised}. Motivated by these findings, we seek to utilize the {domain-invariant knowledge across real and synthetic data}, to assist the downstream models in overcoming the domain discrepancy. 

Unlike the individual speech frames that are sensitive to channel and noise variation, contextual speech flow is a prominent auditory cue for speech perception and cognition across diverse scenarios~\cite{oord2018representation}. According to the studies by Bregman et al.~\cite{bregman1994auditory, irvine2012auditory,guinan2006olivocochlear,szabo2016computational}, the cognition of specific speech signals in a complex environment is guided by the contextual knowledge. For example, in the study of double synthetic vowel experiment~\cite{warren1971speech, warren1972auditory, weintraub1985theory}, psychoacousticians reveal that listeners unconsciously infer the most likely sound through the contextual cues, including phoneme and word information, during the perceptual process, which is referred to as the \textit{auditory induction}~\cite{lee2011single}. In this paper, we are inspired by the prior findings to study a domain-invariant pretrained (DIP) frontend that captures contextual cues from unlabeled, real speech mixtures. With such a DIP frontend, we anticipate that the downstream speech separation models will acquire transferable knowledge for real-world scenarios.

In practice, the proposed DIP frontend is a learnable contextual cue extractor trained on some pretext tasks with large-scale wild unlabeled or weakly labeled data. As demonstrated in~\cite{yue2022self}, there are two typical types of pretext tasks for pretrained frontend, namely generative pretext task and discriminative pretext task. 
The generative pretext task~\cite{chung2020generative,liu2020mockingjay,liu2021tera,chi2021audio,chung2020vector,ling2020deep,chen2023speech} treats the speech itself as a sequential structure and reconstruct the detailed information based on the prior context. On the other hand, the discriminative pretext task~\cite{schneider2019wav2vec, baevski2020wav2vec, wang2022wav2vec,hsu2021hubert,chen2022unispeech,chen2022wavlm,lian2023av,wang2023adapter,fazel2023cocktail,ng2023hubert, lin2024selective,yue2022self,baevski2022data2vec,wang2023data2vec,zhu2023robust, hu2023wav2code} focuses on capturing the high-level ``slow features''~\cite{oord2018representation} that span many time steps, through predicting the masked frames in a contrastive manner or the index of masked clustering labels. 

We advocate that contrastive predictive coding (CPC)~\cite{oord2018representation,schneider2019wav2vec,baevski2020wav2vec,wang2022wav2vec}, a discriminative pretext task, is suitable for reducing the domain mismatch between real and synthetic data because it learns contextual cues by encoding multi-scale information~\cite{pasad2021layer} with mutual information maximization.  In this paper, we adopt a Siamese network as the domain-invariant pretrained frontend, and devise the pretraining strategy. The contributions of our paper are summarized as follows,

\begin{itemize}
    \item We introduce the concept of using  mixture waveform for self-supervised frontend pretraining and design pretext task to capture the contextual cues with both real and synthetic inputs.
    \item We derive the domain discrepancy distance and formulate a Siamese network with a novel domain loss to further reduce domain mismatch.
    \item We propose a general pipeline to incorporate popular frontends into various separation models.
    \item We conduct comprehensive experiments to validate the superior performance of DIP frontend across diverse real and synthetic datasets, with various downstream models, and different evaluation metrics. 
\end{itemize}

The rest of the paper is organized as follows. Section ~\ref{sec:related_work} discusses the prior studies on pretrained frontend. Section ~\ref{sec:model} formulates the pretraining method and introduces the DIP model architecture. In Section ~\ref{sec:exp_setup}, we describe the model configuration and experimental setup. In Section ~\ref{sec:exp_result}, we report the results. Finally, Section ~\ref{sec:con_ssl} concludes the paper.

\section{Related Work on Self-Supervised Pretraining}
\label{sec:related_work}

The objective of frontend pretraining is to capture contextual cues from raw observations through task-agnostic pretext tasks. Generally, the pretext tasks employed for frontend pretraining fall into two categories: generative and discriminative. The former emphasizes learning representations by reconstructing input data, while the latter relies on predicting future or missing frames.

\subsection{Autoregressive Predictive Coding}

 One of the generative methodologies is autoregressive predictive coding (APC), which is inspired by the long-range dependency observed in the raw audio waveform. The APC assumes that the probability distribution of latent features at a specific time step is conditionally dependent on the preceding information in the raw input sequence. In this perspective, the frontend can learn to capture the contextual cues of the audio waveform by minimizing the feature distance between the predicted sequence and the target sequence. 
 
 In~\cite{chung2020generative}, Chung et al. first explore learning speech representation through a multi-layer unidirectional LSTM frontend in a generative manner. By reconstructing the Mel spectrogram of future frames with previous information, the APC  demonstrates significant improvement in the phoneme classification task. The VQ-APC~\cite{chung2020vector} further extends this method through vector-quantization and multi-task training, since downstream tasks are inherently discrete. Additionally,  Decoar~\cite{ling2020deep} leverages both forward LSTM and backward LSTM to equip the frontend with bidirectional information. However, the APC-based pretrained frontend may not be conducive to speech separation, as reported in~\cite{tsai2022superb}. This limitation could be attributed to the fact that APC relies on reconstructing the magnitude of the waveform while disregarding phase information.

\subsection{Masked Acoustic Modeling}
Another generative pretext task is masked acoustic modeling (MAM), which applies the well-performing masked language modeling~\cite{devlin2018bert} pretraining strategy from natural language processing to continuous speech. Unlike the APC which predicts the future frames in an autoregressive manner, MAM forces the frontend to generate a probability distribution of the entire sequence through a fill-in-the-blank task with one or multiple missing elements, technically called the \textit{cloze task}~\cite{taylor1953cloze}. During the pretraining stage, the linear-scale spectrogram of the input waveform is randomly masked or replaced and the frontend is required to reconstruct the details of these frames. The first MAM-based pretrained frontend, Mockingjay~\cite{liu2020mockingjay}, transforms the input waveform into log Mel-features and randomly masks a specific percent of the frames for a multi-layer transformer network to predict. Tera~\cite{liu2021tera} improves the frontend by incorporating both time and frequency masking with various mask ratios. In Data2Vec~\cite{wang2023data2vec, zhu2023robust}, researchers introduce a universal teacher-student learning mechanism that encourages the student model to predict the high-level representation from the teacher model with masked input.   

\subsection{Masked Pseudo-label Prediction}
While the generative pretext tasks typically focus on detailed reconstruction, the discriminative pretext tasks compel the frontend to extract contextual cues through a classification approach. Among all discriminative pretext tasks, the masked pseudo-label prediction task introduced by Hubert~\cite{hsu2021hubert} stands out as a highly favored method for achieving superior results. Given the audio waveforms that are continuous-valued without a prior lexicon, the masked pseudo-label prediction pretext task involves an initial offline clustering step to collect pseudo-labels. Subsequently, the masked sequence is fed into the frontend to predict these labels. In this way, the frontend learns to discover the unit information in a self-supervised manner. 

Recently, plenty of Hubert-based variants have been proposed for speech separation downstream tasks. In Unispeech-sat~\cite{chen2022unispeech}, Chen et al. introduce a primary speaker denoising (PSD) pretext task, enabling the frontend to capture the high-level hidden units from mixture speech. PSD collects the frame-level cluster labels of target reference speech, similar to Hubert, and requires the frontend to predict these labels with synthetic mixture waveform input. WavLM~\cite{chen2022wavlm} extends this framework with relative positional encoding in Transformer architecture and  validates the effectiveness of the frontend in various downstream tasks. To further improve the frontend performance on the speech separation task, Wang et al.~\cite{wang2023adapter,fazel2023cocktail, ng2023hubert, lin2024selective} design a multiple pseudo-label pretext task to capture the hidden unit information of all speakers in the mixture waveform. Despite these variants showing convincing improvement in the downstream speech separation task, the requirement for both mixture waveform and its corresponding target reference speech limits its usage in real-world scenarios. 

\subsection{Contrastive Predictive Coding}
Instead of predicting pseudo-labels as Hubert, contrastive predictive coding (CPC) aims to capture contextual cues through a contrastive learning methodology with noise-contrastive
estimation (NCE)~\cite{mnih2012fast,gutmann2010noise,jozefowicz2016exploring}. The CPC pretext task draws inspiration from the predictive coding theories~\cite{rao1999predictive} of neuroscience, which suggests that our brain has a strong ability to predict observations at various levels of abstraction. To emulate this cognitive capacity, the CPC forces the frontend to encode the sequence into compact contextual cues in a way that maximally preserves the mutual information. 

The contrastive predictive coding for pretrained frontend originates from the research conducted by Oord et al.~\cite{oord2018representation}. The authors employ a convolutional neural network (CNN) to encode the input sequence into latent space and implement a stacked recurrent neural network (RNN) to capture global abstraction with InfoNCE loss. In Wav2vec~\cite{schneider2019wav2vec}, Schneider et al. extend this contrastive learning framework with causal convolution structure, showcasing overwhelming results in both ASR and phoneme recognition tasks. Wav2vec2.0~\cite{baevski2020wav2vec}, introduced by Baevski et al., improves the model by combining the Transformer network with trainable discrete codebooks. In~\cite{pasad2021layer}, Pasad et al. further 
 demonstrates that the frontend with CPC pretext task can capture the abstraction of Mel spectrogram, phonemes, and words via layer-wise analysis. 

Among the four pretraining pretext tasks, the CPC method is particularly suitable for frontend pretraining to narrow the domain gap between real and synthetic mixtures across multiple scales. This is attributed to the close relationship between the mutual information of the mixture waveform and the contextual cues of the target reference speech. In the following section, we demonstrate this insight theoretically and employ contrastive learning to accomplish frontend pretraining.

\section{Modeling Contextual Cues}
\label{sec:model}

There have been attempts to extract contextual cues from speech mixture, e.g. Unispeech-sat~\cite{chen2022unispeech}, WavLM~\cite{chen2022wavlm}, and Cocktail-Hubert~\cite{fazel2023cocktail}, through pretraining. Unfortunately, these techniques depend on a parallel corpus between a speech mixture and its individual voices. Such parallel corpus is unavailable in real-world scenarios. In this paper, we assume that a speech mixture, either real or synthetic, is an additive sum of two individual voices. However, the individual voices are not available during training and testing. 

\begin{figure*}[t]
	\centering 
	\includegraphics[width=1.0\linewidth]{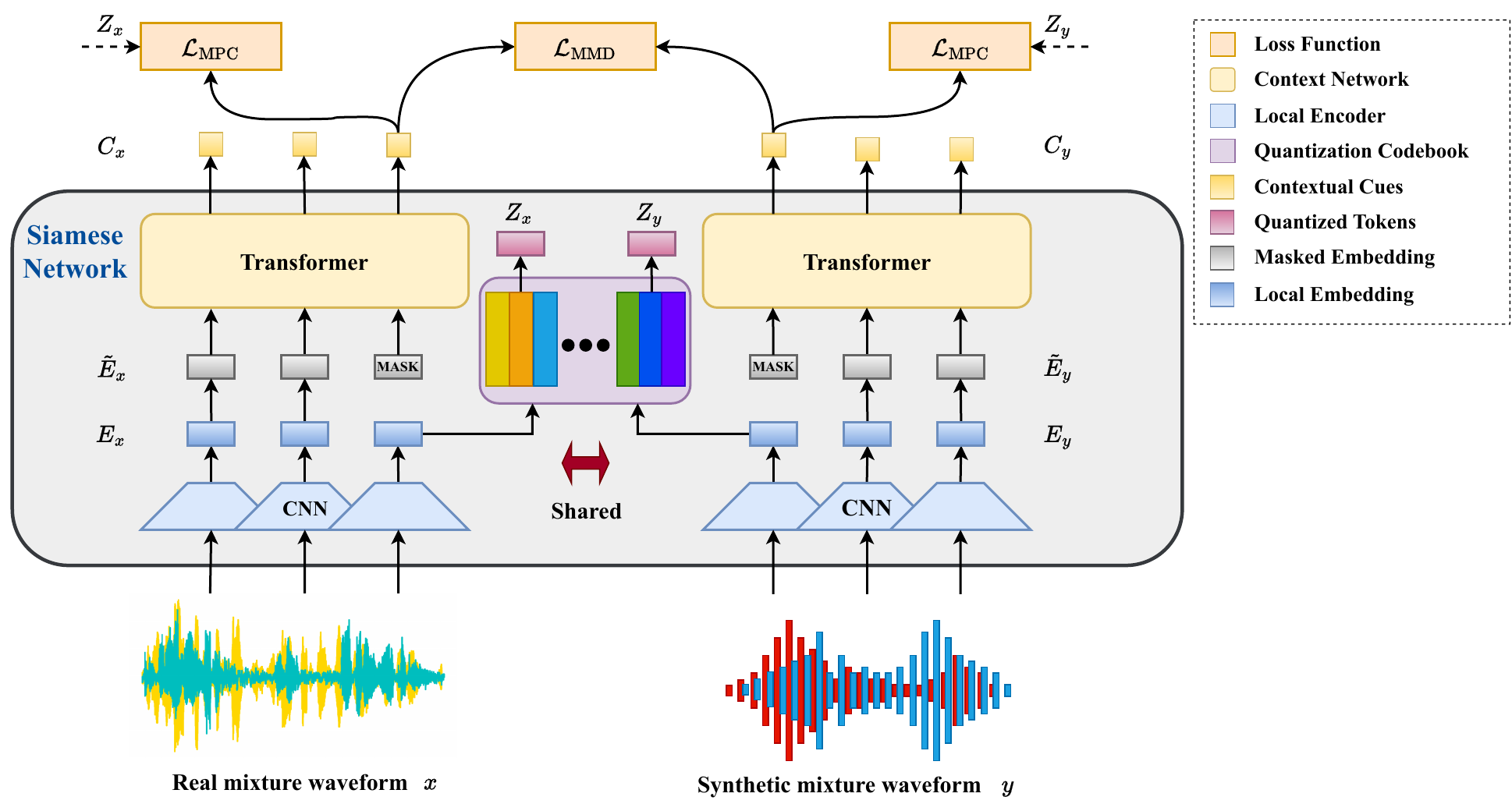}
	\caption{The architecture of our DIP frontend. We pretrain the frontend with both real and synthetic mixtures through a Siamese network to capture the contextual cues. The $\boldsymbol{x}$ and $\boldsymbol{y}$ are randomly sampled mixture waveforms from the real mixtures and synthetic datasets. The $\mathbf{E}_x$ and $\mathbf{E}_y$ are the local embedding extracted by the local encoder $h_{\theta}$. The $\tilde{E}_x$ and $\tilde{E}_y$ are the randomly masked embedding with percentage $\delta$. The $\mathbf{Z}_x$ and $\mathbf{Z}_y$ are the quantized embeding with codebook $q_{\theta}$ for siamses network to predict. The $\mathbf{C}_x$ and $\mathbf{C}_y$ are contextual cues captured by the Transformer $g_{\theta}$. The $\mathcal{L}_{MPC}$ is the contrastive loss defined at Eq.~\eqref{equ.mpc} and $\mathcal{L}_{MMD}$ is the discrepancy loss in Eq.~\eqref{equ.mmd_rep}.} 
	\label{Fig.structure} 
\end{figure*}

We first propose a mixture predictive coding (MPC) pretext task to capture contextual cues from the mixture waveform. We then extend it to a mixture invariant coding (MIC) pretext task with a domain discrepancy loss to further minimize the domain gap between the real and synthetic samples. An overall model structure of the proposed domain-invariant pretrained frontend is depicted in Fig.~\ref{Fig.structure}. 

\subsection{Mixture Predictive Coding}

Mutual information (MI) serves as a widely employed fundamental measure for assessing the non-linear statistical relationship between two random variables. In a previous study~\cite{oord2018representation}, Oord et al. first apply contrastive learning to single-speaker speech pretraining, by maximizing the mutual information between contextual cue and reference speech. Extending the idea to multi-speaker speech, we introduce the mixture predictive coding (MPC) for frontend pretraining with unlabeled multi-speaker mixture input. We expect that the pretrained frontend learns the contextual cues of target speakers in a cocktail party.

Let $\boldsymbol{m} \in \mathbb{R}^{1 \times \tau}$ be a multi-speaker mixture waveform from the unlabeled joint mixture dataset $\mathcal{D}_m =\mathcal{D}_x \cup  \mathcal{D}_y$, where $\mathcal{D}_x$ is the real mixture dataset and $\mathcal{D}_y$ is the synthetic mixture dataset. We assume that all $L$ isolated target reference speech $\boldsymbol{s}_i$ are mixed in an additive way,

\begin{equation}
   \boldsymbol{m} = \sum_{i=1}^L \boldsymbol{s}_i,
\end{equation}
Here $\boldsymbol{s}_i \in \mathbb{R}^{1 \times \tau}$ is the $i$ single-speaker reference speech in the mixture. As shown in CPC~\cite{oord2018representation},  a proper frame-based\footnote{In this paper, a variable with $\tau$ as the index represents a sequence of samples
in the time-domain, while a variable with $T$ as the index represents a sequence
of frame-based embeddings.} contextual cues $\mathbf{C}_m \in \mathbb{R}^{T \times D}$ with $D$ channels should maximize the mutual information $I(\mathbf{C}_m; \boldsymbol{s}_i)$ to encode the information of reference speech $\boldsymbol{s}_i$. Similarly, for the multi-speaker mixture waveform, since all reference speech is independent of each other, we can obtain the contextual cues $\mathbf{C}_m$ by maximizing the mutual information $I(\mathbf{C}_m; \boldsymbol{s}_i)$ of each reference speech $ \boldsymbol{s}_i$. However, due to the absence of reference speech for the real mixture, direct optimization of $I(\mathbf{C}_m; \boldsymbol{s}_i)$ is not feasible. Therefore, we propose an alternative method that maximizes the lower bound of all $I(\mathbf{C}_m; \boldsymbol{s}_i)$.

We treat the reference speech as isolated random variables. Given the assumption of independence, the entropy of the added variables is greater than the maximum entropy among all variables. Consequently, we have the lower bound of the mutual information for all reference speech $\boldsymbol{s}_i$ as follows:
\begin{equation}
    \begin{split}
        I\left(\mathbf{C}_m; \boldsymbol{s}_i\right) & = H\left(\boldsymbol{s}_i\right) - H\left(\boldsymbol{s}_i|\mathbf{C}_m\right)\\
        & \geq \left(H\left(\boldsymbol{s}_i\right) - H\left(\boldsymbol{m}\right)\right) + \left(H\left(\boldsymbol{m}\right) - H\left(\boldsymbol{m}|\mathbf{C}_m\right)\right)\\
        & = \underbrace{\left(H\left(\boldsymbol{s}_i\right)-H\left(\boldsymbol{m}\right)\right)}_{\text{Constant}} + \underbrace{I\left(\mathbf{C}_m;\boldsymbol{m}\right)}_{\text{Mixture MI}},
    \end{split}
    \label{equ.mi}
\end{equation}

The left-hand side of Eq.~\eqref{equ.mi} represents the mutual information of arbitrary reference speech, while the right-hand side is its lower bound. The first component $[H(\boldsymbol{s}_i)-H(\boldsymbol{m})]$ is a constant value denoting the entropy difference between the mixture waveform $\boldsymbol{m}$ and its reference speech $\boldsymbol{s_i}$. The second component $I(\mathbf{C}_m; \boldsymbol{m})$ measures the mutual information between the contextual cues $\mathbf{C}_m$ and the mixture sequence $\boldsymbol{m}$. The Eq.~\eqref{equ.mi} illustrates that \textit{maximizing the $I(\mathbf{C}_m;\boldsymbol{m})$ of mixture waveform is equivalent to maximizing the lower bound of $I(\mathbf{C}_m; \boldsymbol{s}_i)$ of reference speech}, thereby encoding the information of reference speech into the contextual cues $\mathbf{C}_m$. 

In this paper, we follow the contrastive learning algorithm in Wav2Vec2.0~\cite{baevski2020wav2vec}. We alternately use the mixture waveform from either real or synthetic datasets as input and pretrain our DIP frontend through maximizing the mutual information $I(\mathbf{C}_m;\boldsymbol{m})$, to learn shared contextual cues across two different domains. During the pretraining stage, we first randomly select normalized mixture waveform $\boldsymbol{m}$ from the joint mixture dataset $\mathcal{D}_m$. Then, we map $\boldsymbol{m}$ into the local feature $\mathbf{E}_m$ through local encoder $h_\theta$ and utilize quantization codebooks $q_\theta$ to discretize it into tokens $\mathbf{Z}_m$. Next, we randomly mask the local embedding $\mathbf{E}_m$ with a specific ratio $\delta$ to obtain masked sequence $\mathbf{\tilde{E}_m}$. Finally, we utilize a contextual network $g_\theta$ to predict the contextual cues $\mathbf{C}_m$ based on the masked sequence $\mathbf{\tilde{E}_m}$. The optimization of the model is achieved through the NCE method with $\mathcal{L}_{\text{MPC}}$ loss function as follows:
\begin{equation}
    \label{equ.mpc}
    \mathcal{L}_{\text{MPC}}\left(\mathbf{C}_m, \mathbf{Z}_m\right) = \mathcal{L}_d + \mathcal{L}_{\text{NCE}}\left(\mathbf{C}_m, \mathbf{Z}_m\right),
\end{equation}
where $\mathcal{L}_d$ is the diversity loss of the $G$ quantization codebook with $V$ entries to encourage the comprehensive use of all quantized codebook representations as Wav2Vec2.0~\cite{baevski2020wav2vec}: 

\begin{equation}
    \mathcal{L}_{d} = -\log \frac{1}{GV}\sum_{g=1}^G\sum_{v=1}^Vp_{g,v}\log p_{g,v}
\end{equation}
where $p_{g,v}$ is the probability of selecting the $v$ quantized code in the $g$ codebook. The $\mathcal{L}_{\text{NCE}}$ with temperature $\omega$ and similarity function $\psi$ stands for the contrastive loss employed by our DIP frontend to distinguish the positive quantized token $\mathbf{Z}_m$ from a set of quantized
candidates $\mathcal{Q}_Z$,  which includes $\mathbf{Z}_m$ and other negative distractors:
\begin{equation}
    \label{equ.mpc_nce}\mathcal{L}_{\text{NCE}}\left(\mathbf{C}_m, \mathbf{Z}_m\right) = -\log \frac{\exp\left(\psi\left(\mathbf{C}_m, \mathbf{Z}_m\right) / \omega\right)}{\sum_{\tilde{\mathbf{Z}} \in \mathcal{Q}_Z} \exp\left(\psi\left(\mathbf{C}_m, \tilde{\mathbf{Z}}\right) / \omega\right)},
\end{equation}

\subsection{Mixture Invariant Coding}

In the MPC pretext task, we illustrate that maximizing the mutual information of unlabeled mixture forces the frontend to capture the contextual cues. However, the learned contextual cues of different domains may still differ from each other, thereby limiting the domain transfer capacity of our DIP frontend. Neural-based domain adversarial training (DAT)~\cite{ganin2015unsupervised,tzeng2017adversarial, hou2019domain, cheng2020deep}, which learns domain indistinguishable representation through gradient reversal layers, is one solution to address this problem. However, the requirement of a suitable discriminator for variable-length sequences and the difficulty of optimizing the gradient reverse layers often lead to unstable convergence results. Another approach, the statistic-based maximum mean discrepancy (MMD)~\cite{gretton2012kernel},  which directly minimizes the maximum statistical difference of different domains without auxiliary networks, demonstrates convincing improvements in various speech tasks~\cite{zhu2022multi, cheng2020deep}. Inspired by this method, we introduce the mixture invariant coding (MIC) pretext task as an extension of the MPC pretext task, to mitigate the domain gap in extracted contextual cues between real and synthetic samples. 

Let $\mathbf{C}_x$ and $\mathbf{C}_y$ be the contextual cues of the real mixture $\boldsymbol{x}$ from $\mathcal{D}_x$ and synthetic mixture $\boldsymbol{y}$ from $\mathcal{D}_y$, respectively. The MMD distance $d_{\mathcal{H}}$ of these cues is based on the embedding probabilities in Reproducing Kernel Hilbert Space (RKHS) ${\mathcal{H}}$,
\begin{equation}\label{equ.mmd}d_{\mathcal{H}}\left(\mathbf{C}_x,\mathbf{C}_y\right) = \sup_{||f||_\mathcal{H}\leq 1}\mathbb{E}_{p({\mathbf{C}_x})}\left[f(C_x)\right] -\mathbb{E}_{p({\mathbf{C}_y})}\left[f(C_y)\right],
\end{equation}
where $p({\mathbf{C}_x})$ and $p({\mathbf{C}_y})$ represents the probability distribution function of contextual cues and $f$ is the support function where the second order norm in $\mathcal{H}$ space is smaller or equal to 1. $\mathbb{E}$ denotes the expectation over the entire probability distribution space of $p({\mathbf{C}_x})$ and $p({\mathbf{C}_y})$. 

In Eq.~\eqref{equ.mmd}, the definition of MMD illustrates that the calculation of the MMD distance relies on all possible support functions $f$ in the $\mathcal{H}$ space. As it is challenging to traverse all support functions, a common approach is to simplify the MMD distance by expanding it with kernel function $\kappa$ and transforming it into a discrete loss function as in~\cite{gretton2012kernel}:
\begin{equation}
    \begin{split}
    \mathcal{L}_{\text{MMD}}\left(\mathbf{C}_x,\mathbf{C}_y\right) & = \sum_{j=1}^M\sum_{k=1}^Mp\left(\mathbf{C}_{x_j}\right)p\left(\mathbf{C}_{x_k}\right)\kappa\left(\mathbf{C}_{x_j},\mathbf{C}_{x_k}\right) \\ & -2\sum_{j=1}^M\sum_{k=1}^Np\left(\mathbf{C}_{x_j}\right)p\left(\mathbf{C}_{y_k}\right)\kappa\left(\mathbf{C}_{x_j},\mathbf{C}_{y_k}\right) \\ &  +\sum_{j=1}^N\sum_{k=1}^Np\left(\mathbf{C}_{y_j}\right)p\left(\mathbf{C}_{y_k}\right)\kappa\left(\mathbf{C}_{y_j},\mathbf{C}_{y_k}\right), \\
    \end{split}
    \label{equ.full_rep}
\end{equation}
where $j$ and $k$ are the sample index. $M$ and $N$ are the total number of possible cues. The $p\left(\mathbf{C}_{x_j}\right)$, $p\left(\mathbf{C}_{x_k}\right)$, $p\left(\mathbf{C}_{y_j}\right)$ and $p\left(\mathbf{C}_{y_k}\right)$ are contextual cues distribution that remain unknown. Conventional MMD-based methods~\cite{thota2021contrastive, zhu2022multi,aharoni2020unsupervised} tackle the unknown probability distribution issue by replacing them with sample frequency through random sampling, which is referred as the Monte Carlo method. However, since the contextual cues distribution of our DIP frontend is updated after every iteration and does not support random sampling, directly applying the Monte Carlo method in frontend pretraining is not feasible. Therefore, we attempt to estimate the contextual cues distribution with the NCE method. For instance, we extend $p(\mathbf{C}_{x_j})$ as follows:
\begin{equation}
    \begin{split}
    p\left(\mathbf{C}_{x_j}\right) & =\sum_{\boldsymbol{x}}p\left(\mathbf{C}_{x_j} \middle| \boldsymbol{x}\right)p\left(\boldsymbol{x}\right),
    \end{split}
    \label{equ.full_prob}
\end{equation}

The summation $\sum_{\boldsymbol{x}}$ involves calculating the total probability by traversing all possible values of $\boldsymbol{x}$. The $p(\boldsymbol{x})$ represents the fixed sample distribution of mixtures. In the view of large-scale pretraining, since the sample diversity is sufficient, this sample distribution can be approximated with a uniform distribution $p(x)=\frac{1}{M}$, which has been proven to be effective in~\cite{thota2021contrastive,zhu2022multi,aharoni2020unsupervised}. The conditional probability distribution $p(\mathbf{C}_{x_j}|\boldsymbol{x})$ denotes the probability massive function of contextual cues given the input sequences, and we propose to approximate it using the NCE estimation $\hat{p}\left(\mathbf{C}_{x_j} \middle| \boldsymbol{x}\right)$:
\begin{equation}
    \begin{split}
    \hat{p}\left(\mathbf{C}_{x_j} \middle| \boldsymbol{x}\right) & = \frac{\exp\left(\psi\left(\mathbf{C}_{x_j},\mathbf{Z}_x\right)\right)}{\sum_{\Tilde{\mathbf{C}} \in \mathcal{Q}_C}\exp\left(\psi\left(\Tilde{\mathbf{C}},\mathbf{Z}_x\right)\right)}, \\ 
    \end{split}
    \label{equ.cond_prob}
\end{equation}

Here $\mathcal{Q}_C$ is a set of $K+1$ contextual cues candidates, which includes positive cues $\mathbf{C}_{x_i}$ and $K$ distractors. A small value of $K$ leads to the estimated contextual cues distribution degrading to the sample distribution and we explore the effectiveness of $K$ in section~\ref{sec:exp_result}. By bringing the $\hat{p}\left(\mathbf{C}_{x_j} \middle| \boldsymbol{x}\right)$ into Eq.~\eqref{equ.full_prob}, we can derive an approximation of the ground-truth contextual cues probability distribution $p(\mathbf{C}_{x_i})$: 
\begin{equation}
    \begin{split}
    \hat{p}\left(\mathbf{C}_{x_j}\right) & = \frac{1}{M}\sum_x\frac{\exp\left(\psi\left(\mathbf{C}_{x_j},\mathbf{Z}_x\right)\right)}{\sum_{\Tilde{\mathbf{C}} \in \mathcal{Q}_C}\exp\left(\psi\left(\Tilde{\mathbf{C}},\mathbf{Z}_x\right)\right)}, \\ 
    \end{split}
    \label{equ.full_prob_rep}
\end{equation}

Likewise, we can apply this method to estimate $\hat{p}(\mathbf{C}_{x_k})$, $\hat{p}(\mathbf{C}_{y_j})$, and $\hat{p}(\mathbf{C}_{y_k})$ and calculate the MMD loss $\mathcal{L}_{\text{MMD}}$:
\begin{equation}
    \begin{split}
    \mathcal{L}_{\text{MMD}}\left(\mathbf{C}_x,\mathbf{C}_y\right) & = \sum_{j=1}^M\sum_{k=1}^M\hat{p}\left(\mathbf{C}_{x_j}\right)\hat{p}\left(\mathbf{C}_{x_k}\right)\kappa\left(\mathbf{C}_{x_j},\mathbf{C}_{x_k}\right) \\ & -2\sum_{j=1}^M\sum_{k=1}^N\hat{p}\left(\mathbf{C}_{x_j}\right)\hat{p}\left(\mathbf{C}_{y_k}\right)\kappa\left(\mathbf{C}_{x_j},\mathbf{C}_{y_k}\right) \\ &  + \sum_{j=1}^N\sum_{k=1}^N\hat{p}\left(\mathbf{C}_{y_j}\right)\hat{p}\left(\mathbf{C}_{y_k}\right)\kappa\left(\mathbf{C}_{y_j},\mathbf{C}_{y_k}\right), \\
    \end{split}
    \label{equ.mmd_rep}
\end{equation}

The pretraining procedure of MIC is as follows: We first randomly select a batch of normalized real mixtures $\boldsymbol{x}$ from real mixtures dataset $\mathcal{D}_x$ and synthetic mixtures waveform $\boldsymbol{y}$ from synthetic mixture dataset $\mathcal{D}_y$. Then we extract the local embeddings $\mathbf{E}_x$ and $\mathbf{E}_y$ from both real mixture and synthetic mixture using a shared local encoder $h_{\theta}$ and quantize them as $\mathbf{Z}_x$ and $\mathbf{Z}_y$ with shared quantization codebooks $q_{\theta}$. Next, we randomly mask $\delta$ percentage frames of local embeddings and feed them into a shared-weight contextual network $g_{\theta}$ to generate contextual cues $\mathbf{C}_x$ and $\mathbf{C}_y$. Finally, we calculate the multi-task loss $\mathcal{L}_{\text{MIC}}$ to optimize our DIP frontend. The multi-task loss $\mathcal{L}_{\text{MIC}}$ is the summation of MPC losses from both real and synthetic inputs, plus a weighted MMD loss as follows:
\begin{equation}
    \label{equ.full_loss}    \mathcal{L}_{\text{MIC}}=\mathcal{L}_{\text{MPC}}\left(\mathbf{C}_x, \mathbf{Z}_x\right)+\mathcal{L}_{\text{MPC}}\left(\mathbf{C}_y, \mathbf{Z}_y\right)+\mathcal{\alpha}\mathcal{L}_{\text{MMD}}\left(\mathbf{C}_x,\mathbf{C}_y\right),
\end{equation}
where $\alpha$ is the hyper-parameter that regulates the contributions of the domain gap.

\subsection{Siamese Network}

The Siamese network is initially introduced by Bromley and LeCun~\cite{bromley1993signature} to address the signature verification problem. Typically, a Siamese network consists of twin networks with shared weights to learn the invariant information of two distinct inputs. With the development of contrastive learning, the Siamese network has gained widespread acceptance in computer vision for large-scale image pretraining~\cite{he2020momentum,caron2020unsupervised,chen2020simple}. In these models, the two different augmentations of an image are treated as a positive pair, and techniques such as memory banks, momentum, or gradient clipping are employed to generate negative pairs, to learn shared information.

As our objective is to capture domain-invariant contextual cues, we utilize the Siamese network to concurrently extract contextual cues from both real and synthetic mixtures and optimize the network with proposed pretext tasks. Our proposed frontend contains three components: CNN encoder, quantization codebook, and Transformer as Fig.~\ref{Fig.structure}.

\subsubsection{Local Encoder \texorpdfstring{$h_{\theta}$}{h\_theta}}
The local encoder $h_{\theta}$ is a pyramid convolutional neural network (CNN) comprising multiple temporal convolution blocks followed by layer normalization~\cite{ba2016layer} and a GELU activation function~\cite{hendrycks2016gaussian}. This configuration is designed to generate spectrum-like local embeddings."

\subsubsection{Quantization Codebook \texorpdfstring{$q_{\theta}$}{q\_theta}}
The quantization codebook $q_{\theta}$ discretizes the local embedding into a finite set of speech representations via product quantization. Additionally, a Gumbel softmax function is employed to ensure fully differentiable back-propagation during the selection of codebooks.

\subsubsection{Contextual network \texorpdfstring{$g_{\theta}$}{g\_theta}}
The goal of contextual network $g_\theta$
is to predict contextual cues with masked input. It contains multiple transformer layers with positional embedding to estimate the temporal relationship among inputs. The GELU activation layer and layer normalization are also added to accelerate the converged speed. 

\subsection{Integration with Separation Models}
\begin{figure}[t]
    \centering 
    \includegraphics[width=1.0\linewidth]{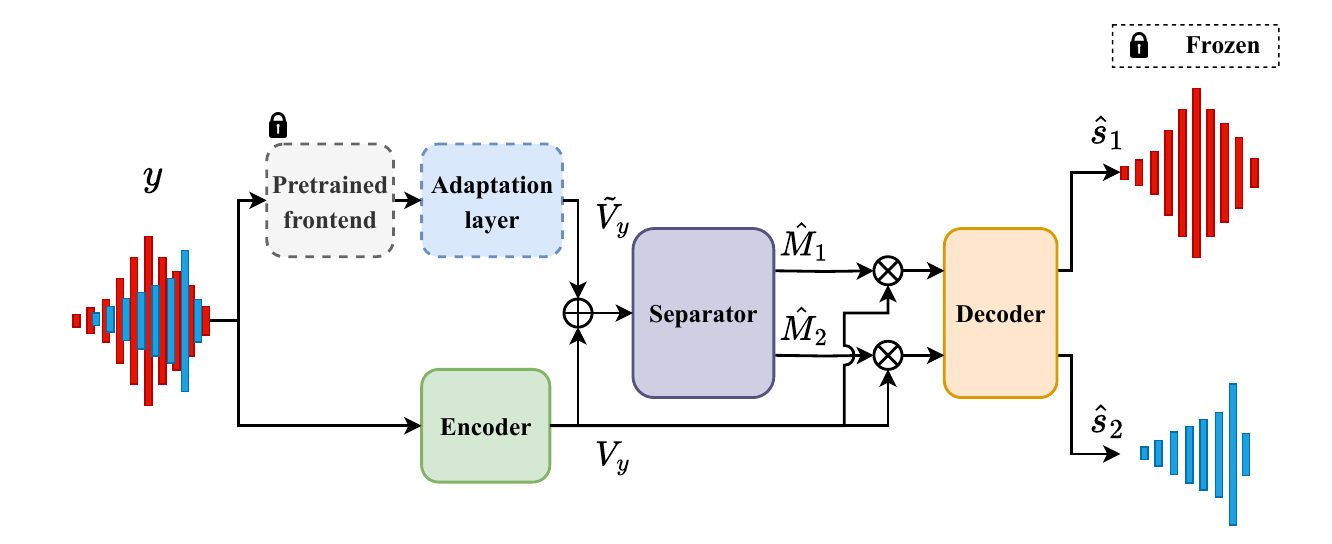}
    \caption{A universal separation training pipeline that incorporates various frontends with diverse speech separation models. A frontend is employed to learn the contextual information, that is followed by an adaptation layer that upsamples the input sequence to time-align with the encoder output for the downstream separation models. The frontend is pre-trained and frozen in the training pipeline. Here $y$ is the two-speaker synthetic mixture waveform. $V_y$ and $\tilde{V_y}$ are the auditory features and contextual cues of mixture waveform. $\hat{M_1}$ and $\hat{M_2}$ are predicted masks for two speakers, respectively, and $\hat{s_1}$, $\hat{s_2}$ are two reconstructed target reference speech.} 
    \label{Fig.pipeline} 
\end{figure}

As mentioned in Section~\ref{sec:introduction}, various separation models operate at different
time resolutions. For example, the feature duration in frequency-domain speech separation models typically ranges from 10 to 20ms, whereas in time-domain separation models, it is generally less than 1ms. Inspired by SUPERB benchmark~\cite{tsai2022superb}, we design a novel pipeline as depicted in Fig.~\ref {Fig.pipeline} to integrate the various frontends into all kinds of conventional speech separation models.

Given a two-speaker mixture waveform $y$, we first feed it into both the encoder and the frontend with an adaptation layer, to generate the auditory embedding $V_y$ and contextual embedding $\tilde{V}_y$. The adaptation layer upsamples the input sequence to time-align with the encoder output. Then we sum them together as input for the separator to predict the mask $\hat{M_1}$ and $\hat{M_2}$ for each speaker. This design maintains the structure of the separator. Finally, we multiply the predicted mask with auditory embedding $V_y$ and reconstruct the target reference speech $\hat{s_1}$ and $\hat{s_2}$ by the decoder.

\section{Experiment Setup}
\label{sec:exp_setup}
\subsection{Experiment Data}
We pretrain our DIP frontend with the mixture waveform first and then train various separators with the labeled synthetic dataset. When evaluating the model's performance, the absence of target reference speech in the real-world mixtures makes it impossible to assess quality using signal-level metrics. Consequently, we divide the experiments of this paper into two parts: a synthetic domain transfer experiment and a real-world scenarios separation experiment.
In the first part, we leverage synthetic datasets to directly evaluate the knowledge transfer capacity of our model through signal-level metrics. For the second part, we assess our model using indirect metrics.

A summary of datasets usage is in Table~\ref{tab_dataset}. For the synthetic domain transfer experiment, we pretrain the frontend with both LM2Mix and Vox2Mix mixture, train the separation model with the LM-train-100 set and test with the test set of all three datasets (LM2Mix, Vox2Mix and LRS2Mix). For the real-world scenarios separation experiment, we do continue pretraining with the REAL-M train set and evaluate the separation quality with its test set.  

\begin{table}[!htb]
	\caption{A summary of datasets usage in our experiments: a ``$\checkmark$'' indicates usage for the current stage, while a ``$-$'' denotes non-usage. The ``Pretrain" indicates usage for the frontend pretraining, ``Train" signifies usage for the separation model training, and ``Test" implies usage for evaluation.}
	\label{tab_dataset}
	\centering 
	\begin{tabular}{|c|c|c|c|c|c|}
	\hline
        \multirow{2}*{Dataset} & Mixture & Reference & \multicolumn{3}{c|}{Data usage}\\ \cline{4-4} \cline{5-5} \cline{6-6}
        ~ & type & type & Pretrain & Train & Test \\
	\hline
        LM2Mix & Synthetic & Anechoic & \checkmark & \checkmark & \checkmark \\
        \hline
        Vox2Mix & Synthetic & Real & \checkmark & - & \checkmark \\
        \hline
        LRS2Mix & Synthetic & Real & - & - & \checkmark \\
        \hline
        REAL-M & Real & N/A & \checkmark & - & \checkmark \\
        \hline
	\end{tabular}
\end{table}

\subsubsection{Synthetic Domain Transfer Experiment}

We employ three synthetic datasets (LM2Mix~\cite{cosentino2020librimix}, Vox2Mix, LRS2Mix~\cite{li2022efficient}) to explore the domain transfer capacity of our DIP frontend in the synthetic domain transfer experiment. Our objective is to transfer the separation knowledge learned from the LM2Mix dataset to the Vox2Mix and LRS2Mix datasets. 

\textbf{LM2Mix}\footnote{\url{https://github.com/JorisCos/LibriMix}}. The dataset is simulated at a 16kHz sampling rate based on Librispeech (LS) corpus~\cite{panayotov2015librispeech}. The simulated dataset is divided into four sets: LM-train-360 set (50,800 utterances, 921 speakers), LM-train-100 set (13,900 utterances, 251 speakers), development set (3,000 utterances, 40 speakers), and test set (3,000 utterances, 40 speakers). All utterances are randomly selected from the LibriSpeech corpus and mixed by specific loudness units relative to full scale. The total hours of LM-train-360, LM-train-100, development, and test are 212h, 58h, 11h, and 11h respectively. 

\textbf{Vox2Mix}. We simulate a two-speakers speech separation dataset at a 16kHz sampling rate based on the Voxceleb2 (Vox2) corpus~\cite{chung2018voxceleb2}. In contrast to the LM2Mix dataset, Vox2Mix dataset incorporates a substantial amount of in-the-wild corpus with different accents, nationalities, attenuation, noise, and reverberation interference. This makes it more challenging and closer to real-world scenarios. The simulated database is partitioned into four sets: Vox-train-253 set (70,000 utterances), Vox-train-53 set (20,000 utterances), development set (3,000 utterances), and test set (3,000 utterances). For the Vox-train-253 set, Vox-train-53 set, and development set, the utterances are randomly selected and mixed from the VoxCeleb2 ``dev'' subset by specific loudness units that are similar to LM2Mix.  Utterances of the test set are generated by ``test'' subset that is not involved in the train set. The total hours of train-253, train-53, development, and test are 253h, 53h, 10h, and 10h, respectively.

\textbf{LRS2Mix}. To assess the generalization capacity of our model, we utilize the test set of LRS2Mix dataset as the unseen set during the test stage. The test set of LRS2Mix contained 3,000  utterances that are simulated from in-the-wild LRS2 dataset~\cite{afouras2018deep}, the spoken sentences audio recordings from BBC television. The two-speaker mixture is simulated from different speaker audios with signal-to-noise ratios sampled between -5 dB and 5 dB. All samples are in 16 kHz and have not been involved in our pretrain and train stages. 

\subsubsection{real-world scenarios Separation Experiment}

For the real-world scenarios separation experiment, we implement our model to the real two-speaker mixture recordings from the REAL-M dataset~\cite{subakan2022real}. This allows us to investigate whether mitigating the domain gap between real and synthetic mixtures contributes to the performance of the separation model in real-world scenarios. 

\textbf{REAL-M}\footnote{\url{https://sourceseparationresearch.com/static/
REAL-M-v0.1.0.tar.gz}} . The samples in the REAL-M dataset are collected by asking contributors to read a predefined set of sentences simultaneously in various acoustic environments with different recording devices such as laptops and smartphones. To facilitate domain-invariant pretraining of our DIP frontend, we partition the entire set of 1436 utterances in the REAL-M dataset into a pretrain set and a test set. The pretrain set comprises 600 mixture samples from the ``early collection'' subset, while the test set contains 837 samples that are not included in the pretrain set. It should be mentioned that all samples from the REAL-M dataset are real recordings without target reference speech. 

\newcolumntype{Y}{>{\centering\arraybackslash}X}
\newcolumntype{Z}[1]{>{\centering\arraybackslash}p{#1}}
\begin{table*}
    \centering
    \caption{A comparison among the ConvTasNet with our DIP frontend, and its variants.  $\alpha$ is the scaling factor in the multi-task loss function in Eq.~\eqref{equ.full_loss}. $K$ is the number of negative distractors in the estimation of feature distribution in Eq.~\eqref{equ.cond_prob}. The development set(Dev) is from LM2Mix. The test sets are from both the matched test set LM2Mix and the mismatched test set Vox2Mix. ``N/A'' in the following tables means no pretrained frontend is used. The MPC is a special case of MIC where $\alpha=0$ as shown in ~\eqref{equ.full_loss}. The \textbf{bold values} are the optimal choices.}
    \begin{tabularx}{\textwidth}{@{}Y|Z{1.2cm}|Z{1.2cm}|Z{1.2cm}|Y|Y|Y|Y|Y@{}}
       \toprule
        \multirow{2}*{System (Sys.)} & Pretrain & \multirow{2}*{$\alpha$} & \multirow{2}*{$K$} & Dev. (LM2Mix) & \multicolumn{2}{c|}{Test (LM2Mix)} & \multicolumn{2}{c}{Test (LM2Mix $\rightarrow$ Vox2Mix)} \\
	 ~ & Strategy & ~ & ~ & SI-SDRi(dB) $\uparrow$ & SI-SDRi(dB) $\uparrow$ & SDRi(dB) $\uparrow$ & SI-SDRi(dB) $\uparrow$ & SDRi(dB) $\uparrow$ \\
  
        \midrule
        baseline & N/A & N/A & N/A & 14.54 & 14.06 & 14.46 & 9.24 & 9.58\\
        
        \midrule 
        1 & MPC & 0 & 0 & 15.58 & 15.30 & 15.69 & 10.62 & 10.94\\

        \midrule 
        2 & \multirow{4}*{MIC} & 100 & 100 & 15.65 & 15.44 & 15.80 & 10.87 & 11.18\\
        \textbf{3} & ~ & \textbf{10} & \textbf{100} &  \textbf{15.96} & \textbf{15.77} & \textbf{16.14}& \textbf{11.05} & \textbf{11.36}\\
        4 & ~ & 1 & 100 & 15.78 & 15.52 & 15.90 & 10.83 & 11.14\\
        5 & ~ & 0.1 & 100 & 15.83 & 15.55 & 15.93 & 10.78 & 11.09\\
        
        \midrule  
        6 & \multirow{4}*{MIC} & 10 & 500 & 15.57 & 15.27 & 15.64 & 10.68 & 11.00\\
        7 & ~ & 10 & 200 & 15.89 & 15.66 & 16.04 & 10.84 & 11.15\\
        8 & ~ & 10 & 20 & 15.95 & 15.76 & 16.13 & 11.02 & 11.33\\
        9 & ~ & 10 & 10 & 15.55 & 15.36 & 15.73 & 10.85 & 11.17\\ 
        10 & ~ & 10 & 0 & 15.49 & 15.23 & 15.60 & 10.77 & 11.09\\
        
       \bottomrule
    \end{tabularx}
    \label{tab_conv1}
\end{table*}

\subsection{Model Configuration}
Our DIP frontend is implemented by fairseq toolkit. The local encoder contains 7 CNN blocks. Each block has 512
channels with strides (5,2,2,2,2,2,2) and kernel widths (10,3,3,3,3,2,2). The convolutional relative positional embedding layer
has kernel sizes of 128 and is divided into 16 groups. For the quantization codebook, we use 2 codebooks. Each codebook has 320 discrete tokens. The Gumbel softmax temperature is annealed from 2 to a minimum of 0.5 by a factor of 0.999995 at every update. The contextual network consists of 12 transformer blocks with a model dimension of 768, an inner dimension of 3,072, and utilizes 8 attention heads. The similarity function $\psi$ is the cosine similarity.

For the downstream separators, we employ three types of time-domain separation models ~\footnote{Note that we present the comprehensive results on ConvTasNet and several key experiments on the others to show the generalizability across different seperation models}(ConvTasNet~\cite{luo2019conv}, DPRNN~\cite{luo2020dual}, and Sepformer~\cite{subakan2021attention}), along with a frequency-domain model(Bi-LSTM~\cite{tsai2022superb}). The implementation of ConvTasNet and DPRNN are based on the Asteroid toolkit, while for Sepformer, we integrate the official implementation from the SpeechBrain toolkit into the Asteroid toolkit to ensure a fair cross-model comparison. All time-domain separation models utilize a 1-D conv layer as the encoder and a 1-D conv-transpose layer as the decoder. The stride and the kernel size of both encoder and decoder are 16 and 32, respectively. Other parameters are similar to the official configuration in the asteroid and speechbrain toolkits. As for the frequency model, we leverage the three-layer Bi-LSTM network in the SUPERB benchmark to estimate the phase-sensitive mask (PSM). The implementation and the checkpoint are available on github\footnote{https://github.com/Wufan0Willan/Mix2Vec}.

\subsection{Training Strategy}
The overall training for our models is conducted in two stages: the pretrain stage and the train stage. 

\subsubsection{Pretrain stage}
The strategy to pretrain our DIP frontend is similar to Wav2vec2.0~\cite{baevski2020wav2vec} in fairseq. We randomly select 65\% frames of all time-steps to be
starting indices and mask the subsequent 10 time steps. Samples in each batch are cropped to the duration of $15.6$ seconds. We also use dropout $0.1$ in the Transformer, at the output of the feature encoder
and the input to the quantization module. In addition, we implement layer drop for the transformer layers with a rate of $0.05$. The learning rate during pretraining stage is $0.0005$ and the pretraining stops when the best validation loss does not improve for $50$ consecutive epochs. We select the Adam with $0.01$ weight decay and $32,000$ warmup steps to optimize the model in the pretraining stage. The
temperature in the contrastive loss is set to $0.1$.

\subsubsection{Train stage}
To verify the compatibility of our DIP frontend, we integrate the contextual cues with both time-domain and frequency-domain separation models. We train all time-domain separation models with recipes in the asteroid toolkit. The batch size of all separation models is 2 and the maximum training epochs is 200. The learning rate is initialized as 1e-3 and decays half when the loss on the validation set is not improved in 5 consecutive epochs. Early stopping is applied if no better model is found in the validation set for 30 consecutive epochs. We utilize Adam optimizer for back-propagation of the separation model only and freeze our DIP frontend during the train stage. For frequency separation models, we follow the standard recipe in the SUPERB benchmark for speech separation tasks. 

\subsection{Evaluation Metrics}

For the synthetic domain transfer experiment, we employ the scale-invariant signal-to-noise ratio improvement (SI-SDRi),  signal-to-noise ratio improvement (SDRi), perceptual evaluation of speech quality (PESQ) and Short Term Objective Intelligibility (STOI) to evaluate the quality of separated speech. As for the real-world scenarios experiment, where target reference speech is unavailable for speech quality evaluation metrics, we select the word error rate (WER) of ASR systems, the Deep Noise Suppression Mean Opinion Score (DNSMOS)~\cite{reddy2021dnsmos} and the neural SNR estimator~\cite{subakan2022real} as the measurement metrics.

\section{Results and Analysis}
\label{sec:exp_result}
\subsection{Effectiveness of Pretrained Frontend}
 
We first validate the effectiveness of our DIP frontend on synthetic datasets through conventional signal distortion metrics. The results are illustrated in Table~\ref{tab_conv1}, where all experiments adopt the ConvTasNet~\cite{luo2019conv} as the separator. We employ the LM2Mix dataset as the source domain set and Vox2Mix as the target domain set. A notable domain shift between these datasets lies in the nature of the target reference speech: Vox2Mix contains real-life reference speech with attenuation, reverberation, and noise, whereas LM2Mix consists of anechoic reference speech. 
We anticipate that the proposed DIP frontend has the capability to alleviate the performance degradation caused by such a domain gap.

\newcolumntype{Y}{>{\centering\arraybackslash}X}
\newcolumntype{Z}[1]{>{\centering\arraybackslash}p{#1}}
\begin{table*}
    \centering
    \caption{An ablation study of pretraining corpus size by the ConvTasNet with our DIP frontend. We utilize the different sizes of pretrain corpus to pretrain the frontend and test the domain transfer capacity from the source domain LM2Mix dataset to the target domain Vox2Mix test set. All pretrain corpus are mixture waveform from synthesis datasets as described in Section \ref{sec:exp_setup}. The \textbf{bold values} are the optimal choices.}
    \begin{tabularx}{\textwidth}{@{}Y|Z{2.2cm}|Z{2.2cm}|Z{2.0cm}|Y|Y|Y|Y@{}}
       \toprule
        System & \multicolumn{2}{c|}{Pretrain Corpus} & Dev. (LM2Mix) & \multicolumn{2}{c|}{Test (LM2Mix)} & \multicolumn{2}{c}{Test (LM2Mix $\rightarrow$ Vox2Mix)}\\ 
		(Sys.) & Source & Target & SI-SDRi(dB) $\uparrow$ & SI-SDRi(dB) $\uparrow$ & SDRi(dB) $\uparrow$ & SI-SDRi(dB) $\uparrow$ & SDRi(dB) $\uparrow$\\
  
        \midrule
        3 & LM2Mix-train100 & Vox2Mix-train53 & 15.95 & 15.77 & 16.14 & 11.05 & 11.36 \\
        11 & LM2Mix-train360 & Vox2Mix-train53 & 16.06 & 15.81 & 16.21 & 11.11 & 11.42 \\
        12 & LM2Mix-train100 & Vox2Mix-train253 & 16.04 & 15.84 & 16.22 & 11.24 & 11.55\\
        \textbf{13} & \textbf{LM2Mix-train360} & \textbf{Vox2Mix-train253} & \textbf{16.55} & \textbf{16.29} & \textbf{16.66} & \textbf{11.71} & \textbf{12.02} \\

       \bottomrule
    \end{tabularx}
    \label{tab_corpus}
\end{table*}

The baseline system outlined in Table ~\ref{tab_conv1} represents the ConvTasNet without any pretrained frontend. This model achieves a performance of 14.06 dB SI-SDRi and 14.46 dB SDRi on the LM2Mix test set, indicating successful application when there is no domain mismatch between the train stage and the test stage. However, directly applying the model to the mismatched Vox2Mix test set (LM2Mix $\rightarrow$ Vox2Mix) without any fine-tuning only achieves 9.24 dB in SI-SDRi and 9.58 dB in SDRi. This result shows that even though both LM2Mix and Vox2Mix datasets are synthetic, the domain gap problem will still significantly damage the separation quality. We can expect that the domain gap will be more serious when transferring the separation knowledge from synthetic data to real samples. 

With $\alpha=0$ (system 1), we implement the MPC pretext task to pretrain the frontend and integrate it with the separation model. The model gains 15.30 dB SI-SDRi and 15.69 dB SDRi for the LM2Mix test set, while for the Vox2Mix test set, the separated speech quality is 10.62 dB in SI-SDRi and 10.94 dB in SDRi. Compared with the baseline system, our DIP frontend model with MPC pretext task improves about 1.24 dB SI-SDRi and 1.23 dB SDRi under the matched condition. For the mismatch condition, it also improves by 1.38 dB SI-SDRi and 1.36 dB SDRi. These results show that by incorporating the contextual cues from our DIP frontend, the downstream model improves the separation quality in both matched and mismatched conditions, which is consistent with the CASA theory as in Section~\ref{sec:introduction}. 
Next, we examine the impact of domain-invariant pretraining with the MIC pretext task. The hyper-parameter $\alpha$ controls the trade-off between MPC loss and MMD loss as in Eq.~\eqref{equ.full_loss}. In general, the larger $\alpha$ forces the pretrained model to pay more attention to alleviating the domain gap, while smaller $\alpha$ means the model should focus on capturing the accurate contextual cues from the mixture. Among systems 2-5, we observe that the system with $\alpha=10$ (system 3) works best on both the LM2Mix and the Vox2Mix test sets. The model achieves 15.77 dB SI-SDRi and 16.14 dB SDRi on the LM2Mix test set, as well as 11.05 dB SI-SDRi and 11.36 dB SDRi on the Vox2Mix test set. These results suggest that with the proposed domain loss, the capability to extract contextual cues from real samples helps our DIP frontend to capture contextual cues from synthetic datasets, and thus further bring about 0.47 dB SI-SDRi and 0.45 SDRi quality improvement when compared with system 1 under matched condition. Similarly, learning to encode contextual cues of synthetic mixtures is also beneficial for our DIP frontend to obtain proper contextual cues on real scenarios and leads to an additional improvement of 0.43 dB SI-SDRi and 0.48 SDRi on mismatched condition. By reducing the $\alpha$ from 10 to 0.1 (system 3-5), we find the model performance on the Vox2Mix test set decline gradually. This observation indicates the significance of mitigating domain gap for effective knowledge transfer in the pretrained frontend. However, if the $\alpha$ is too large, as seen in system 2, the test results of the Vox2Mix test set remains competitive, but the performance on the LM2Mix test set drops a lot. It might be due to the joint optimization of multi-task loss. The large $\alpha$ makes the frontend concentrate on the optimization of eliminating domain gap and ignores the discovery of contextual cues. 

We then proceed to evaluate the model with $K$ number of negative samples, which impacts the estimation of the conditional probability in Eq.(~\ref{equ.cond_prob}). Larger $K$ implies a stronger belief in the normal distribution conditional probability, while small $K$ suggests a tighter, unimodal distribution. Comparing the results between system 3 and systems 6-10, we identify the optimal hyper-parameter value $K=100$ (system 3), indicating an unimodal distribution with a large variance for the true conditional probability. Additionally, We observe that an accurate conditional probability assumption of contextual cues is beneficial for the model to generalize in both the source domain and the target domain, while an 
inappropriate probability assumption will significantly undermines the domain transfer capacity. 

\newcolumntype{Y}{>{\centering\arraybackslash}X}
\newcolumntype{Z}[1]{>{\centering\arraybackslash}p{#1}}
\begin{table*}
    \centering
    \caption{The LM2Mix $\rightarrow$ Vox2Mix domain transfer experiment results among the ConvTasNet separation model. We transfer the separation knowledge from the LM2Mix dataset to the Vox2Mix test set with state-of-the-art pretrained frontends. The Mix corpus~\cite{chen2022unispeech} is an enriched dataset that includes LibriSpeech, GigaSpeech, and VoxPopuli datasets. The UM is the utterance-mixing~\cite{chen2022wavlm} method to generate the mixture waveform in a dynamic way as input. The Mixture-500h pretrain corpus is the mixture waveform of LM2Mix and Vox2Mix we use in system 13. The \textbf{bold values} are the optimal choices.}
    \addtolength{\tabcolsep}{-3.5pt}
    \resizebox{0.95\linewidth}{!}{
    \begin{tabularx}{\textwidth}{@{}Y|Z{2.5cm}|Z{3.0cm}|Y|Y|Y|Y|Y@{}} 
       \toprule
        \multirow{2}*{Separator} & \multirow{2}*{Pretrain model} & Pretrain & Val. (LM2Mix) & \multicolumn{2}{c|}{Test (LM2Mix)} & \multicolumn{2}{c}{Test (LM2Mix $\rightarrow$ Vox2Mix)}\\
		~ & ~ & Corpus & SI-SDRi(dB) $\uparrow$ & SI-SDRi(dB) $\uparrow$ & SDRi(dB) $\uparrow$ & SI-SDRi(dB) $\uparrow$ & SDRi(dB) $\uparrow$\\
        
        \midrule
        \multirow{8}*{ConvTasNet} & N/A & N/A & 14.54 & 14.06 & 14.46 & 9.24 & 9.58 \\
         
        ~ & Wav2vec2.0 & LS 960h & 14.35 & 14.06 & 14.42 & 9.23 & 9.56\\
        
        ~ & Hubert & LS 960h & 13.49 & 14.15 & 14.53 & 9.50 & 9.84\\
        ~ & Unispeech-sat & LS 960h + UM & 15.50 & 15.30 & 15.67 & 10.40 & 10.72\\
        ~ & WavLM & LS 960h + UM & 15.60 & 15.37 & 15.74 & 10.30 & 10.63\\
         ~ & Unispeech-sat+ & Mix 94,000h + UM & 15.22 & 15.04 & 15.39 & 10.44 & 10.76\\
        ~ & WavLM+ & Mix 94,000h + UM & 15.13 & 14.90 & 15.27 & 10.36 & 10.67\\
        ~ & \textbf{DIP} & \textbf{ Mixture 500h} & \textbf{16.55} & \textbf{16.29} & \textbf{16.66} & \textbf{11.71} & \textbf{12.02}\\
       \bottomrule
    \end{tabularx}
    }
    \addtolength{\tabcolsep}{3.5pt}
    \label{tab_conv3}
\end{table*}

\vspace*{-1mm}
\subsection{Impact of the Pretrain Corpus}

The goal of self-supervised pretraining is to acquire high-level contextual cues using massive unlabeled data. Generally, a larger pretraining dataset is expected to be beneficial for the frontend to improve the quality of the downstream model~\cite{hsu2021robust}. Therefore, the corpus size of the unlabeled dataset is crucial for the frontend pretraining. In table ~\ref{tab_corpus}, we investigate the impact of pretrain corpus size. 

We first expand the source corpus from LM2Mix-100 to LM2Mix-360 as system 11. The separation model performance slightly improves by about 0.06 dB SI-SDRi and 0.08 dB SDRi. In the case of the target domain corpus, we replace the Vox2Mix-53 target domain corpus with the Vox2Mix-train253 as system 12. The downstream model gains 0.19 dB quality improvement in both SI-SDRi and SDRi. These results reveal that increasing the corpus size of either the source domain or the target domain can not significantly affect the separation quality. When both the source domain and the target domain corpus are simultaneously enlarged, as in system 13, the model gets significant improvement. The SI-SDRi and SDRi increase by 0.66 dB and 0.68 dB, respectively. These results clearly demonstrate that the frontend pretrained by a large pretraining corpus is helpful for the downstream model to improve the domain transfer capacity.

\begin{figure}[t]
	\centering 
\includegraphics[width=0.85\linewidth]{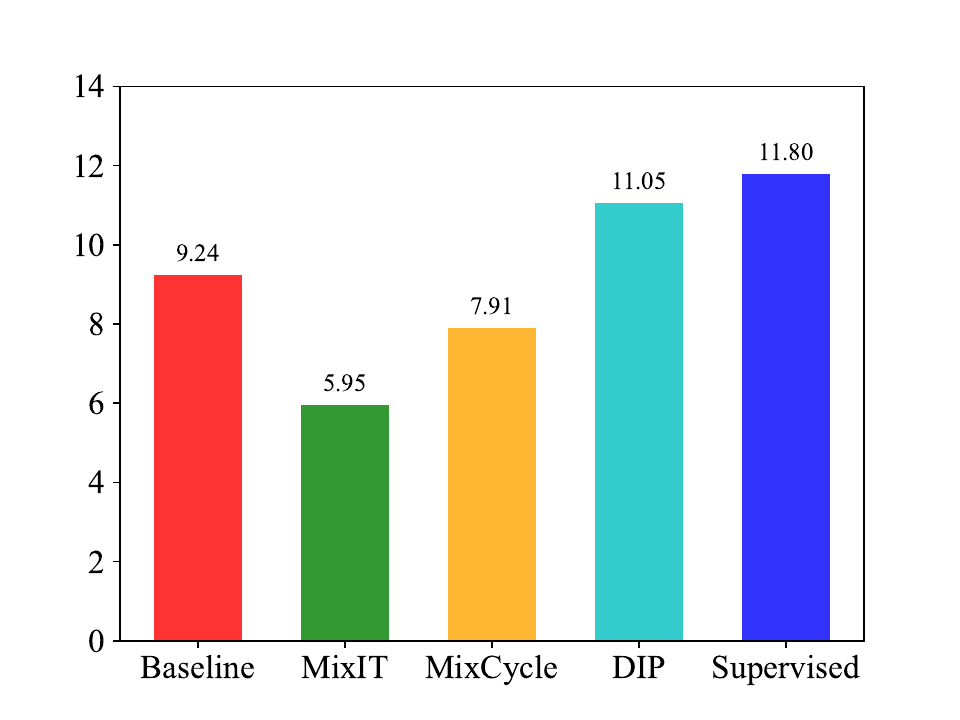}
	\caption{The SI-SDRi(dB) of different methods to separate mixture without from the Vox2Mix test set access to the clean reference speech. The ``Baseline'' is the ConvTasNet without any frontend, the ``DIP'' is the ConvTasNet with our DIP frontend, and the ``Supervised'' is the ConvTasNet trained on Vox2Mix dataset without any frontend.} 
	\label{Fig.unsupervised} 
\end{figure}

\newcolumntype{Y}{>{\centering\arraybackslash}X}
\newcolumntype{Z}[1]{>{\centering\arraybackslash}p{#1}}
\begin{table}
    \centering
    \caption{Experiment results to transfer the separation knowledge from the LM2Mix dataset to the LRS2Mix test set with the ConvTasNet separator. The LRS2Mix test set never appears in both pretrain and train stages. ``$^{\ast}$'' means the results are reported in the original paper. The \textbf{bold values} are the optimal choices.}
    \begin{tabular}{c|c|c|c} 
       \toprule
        \multirow{2}*{Separator} & \multirow{2}*{Frontend} & \multicolumn{2}{c}{Test (LM2Mix $\rightarrow$ LRS2Mix)} \\
	~  & ~  & SI-SDRi(dB) $\uparrow$ & SDRi(dB) $\uparrow$ \\  
        \midrule 
        Supervised~\cite{li2022efficient} & N/A & 10.6$^{\ast}$ & 11.0$^{\ast}$\\
        
        \midrule    
        \multirow{8}*{ConvTasNet} & N/A & 8.62 & 9.10\\
        ~ & Wav2Vec2.0 & 9.11 & 9.55\\
        ~ & Hubert & 9.07 & 9.51\\
        ~ & Unispeech-sat & 10.17 & 10.57\\
        ~ & WavLM & 10.21 & 10.61\\
        ~ & Unispeech-sat+ & 10.52 & 10.92\\
        ~ & WavLM+ & 10.51 & 10.90\\ 
        
        ~ & \textbf{DIP} & 	\textbf{11.99} & \textbf{12.35}\\
       \bottomrule
    \end{tabular}
    \label{tab_lrs}
\end{table}

\subsection{Separation Quality against Other Methods}
To investigate the separation quality against other methods, we first turn our attention to evaluating the downstream model in Table~\ref{tab_conv3} using various pretrained frontends. Compared with the baseline model, the Wav2Vec2.0 and Hubert pretrained frontend yield limited improvements in separation quality. It might becuase these two frontends are pretrained by single-speaker reference speech, making it difficult to predict the contextual cues required by downstream models. The Unispeech-sat and WavLM, pretrained with the primary speaker denoising task~\cite{chen2022wavlm}, effectively capture contextual cues of the primary speaker from mixture waveforms. These cues assist the separation model to improve the speech quality on both the matched LM2Mix test set and unmatched the Vox2Mix test set. However, the Unispeech-sat+ and WavLM+ that pretrained by large-scale corpus fail to bring about significant improvements. In contrast, our proposed DIP frontend, pretrained through the MIC pretext task with ``Mixture 500h'' corpus\footnote{The Unispeech-sat+ and WavLM+ utilize the UM to synthesize mixture waveform dynamically during pretraining. In this view, the ``Mixture 500h'' pretrain corpus we used is equivalent to a subset of their pretrain corpus.}, captures contextual cues of all speakers across different domains and outperforms other pretrained frontends on both LM2Mix and Vox2Mix test sets. Compared to the separation model without a pretrained frontend, our model brings about 2.47 dB SI-SDRi and 2.44 dB SDRi separation quality improvement on the Vox2Mix test set, indicating that our DIP frontend effectively helps the separation model to transfer the separation knowledge from LM2Mix dataset to the Vox2Mix test set.

We further compare the separation quality of our proposed method with other state-of-the-art unsupervised methods using their reference settings\footnote{https://github.com/ertug/MixCycle/tree/main}. We assume that only the mixture waveform of the Vox2Mix-train53 set is available and test the separation models on the Vox2Mix test set. As shown in Fig.~\ref{Fig.unsupervised}, the MixIT and MixCycle achieve 5.95 dB and 7.91 dB SI-SDRi improvement, respectively. The separation quality is even worse than that applying the ConvTasNet ``Baseline'' (trained on the LM2Mix) directly to the Vox2Mix test set. This could be a result of target mismatch, in view of the fact that these methods utilize the mixture waveform as the target instead of a clean reference, as described in Section.~\ref{sec:introduction}. The ConvTasNet ``DIP'' (Sys.3) trained on the LM2Mix dataset with our proposed DIP frontend obtains an 11.05 dB SI-SDRi, almost the same as the supervised separation model trained on the Vox2Mix training set. This result demonstrates the strong domain transfer capacity of our method. 

\begin{figure}[t]
	\centering 
\includegraphics[width=0.85\linewidth]{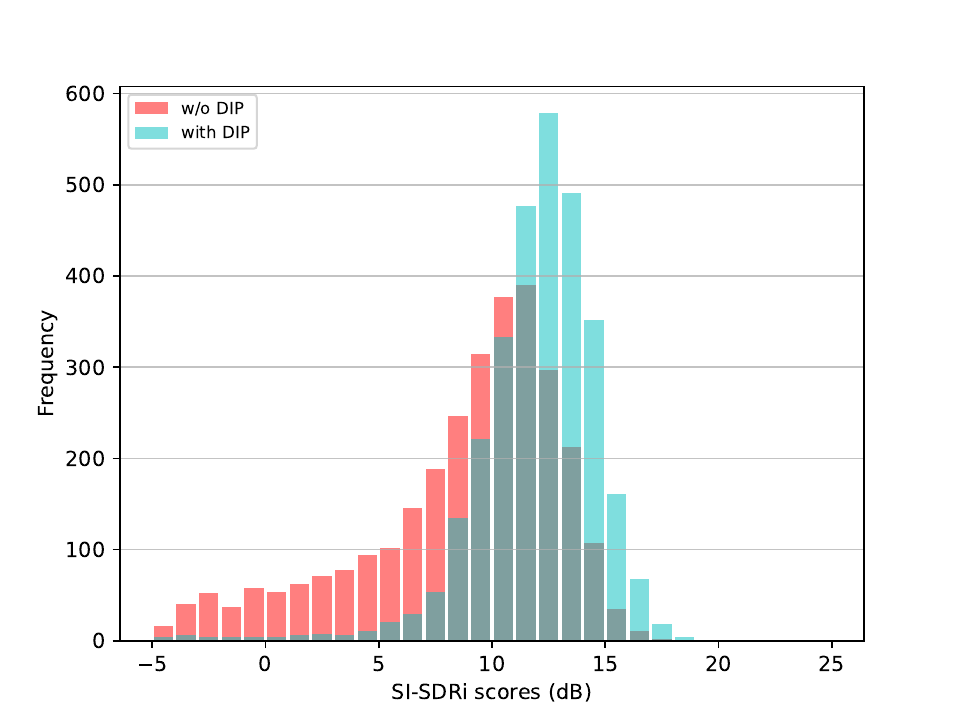}
	\caption{The SI-SDRi(dB) score of the scores on individual utterances. The red bar is the number of samples for special SI-SDRi(dB) using the ConvTasNet baseline model. The blue bar on the ConvTasNet with our DIP frontend.} 
	\label{Fig.LRS2Mix} 
\end{figure}

\subsection{Generalization Ability on Unseen Dataset}

To assess the robustness of our proposed DIP frontend on out-of-domain scenarios, we employ the test set of LRS2Mix dataset (LM2Mix $\rightarrow$ LRS2Mix) and evaluate various frontends without fine-tuning. The target reference speech in the LRS2Mix test set is real recordings collected from BBC television. As illustrated in Table~\ref{tab_lrs}, the performance of the baseline model significantly drops to 8.62 dB SI-SDRi and 9.10 dB SDRi when compared to the supervised ConvTasNet model trained on the matched LRS2Mix training set (10.6 dB SI-SDRi and 11.0 dB SDRi), due to the domain gap between LRS2Mix and LM2Mix. With the pretrained frontend, the model performance on the LRS2Mix dataset improves. Our proposed DIP frontend achieves outstanding results with SI-SDRi and SDRi values of 11.99 dB and 12.35 dB, respectively, surpassing all other frontends. These results demonstrate the effectiveness of our DIP frontend to improve the domain transfer capacity, even for unseen datasets.

In Fig.~\ref{Fig.LRS2Mix}, we investigate the SI-SDRi improvement of individual utterances. The red bars represent samples from the ConvTasNet baseline without any frontend. Most of these samples are above 5 dB, while many are below 0 dB. This indicates that, due to the domain-mismatch problem, the signal distortion of separated speech may be worse than the original mixture. The blue bar is the samples from the ConvTasNet with our DIP frontend. By utilizing the frontend, the sample SI-SDRi improves significantly. Most samples are above 10 dB and almost all samples are above 0 dB. These results also show the effectiveness of our DIP frontend.

\begin{figure}[t]
	\centering 
\includegraphics[width=0.85\linewidth]{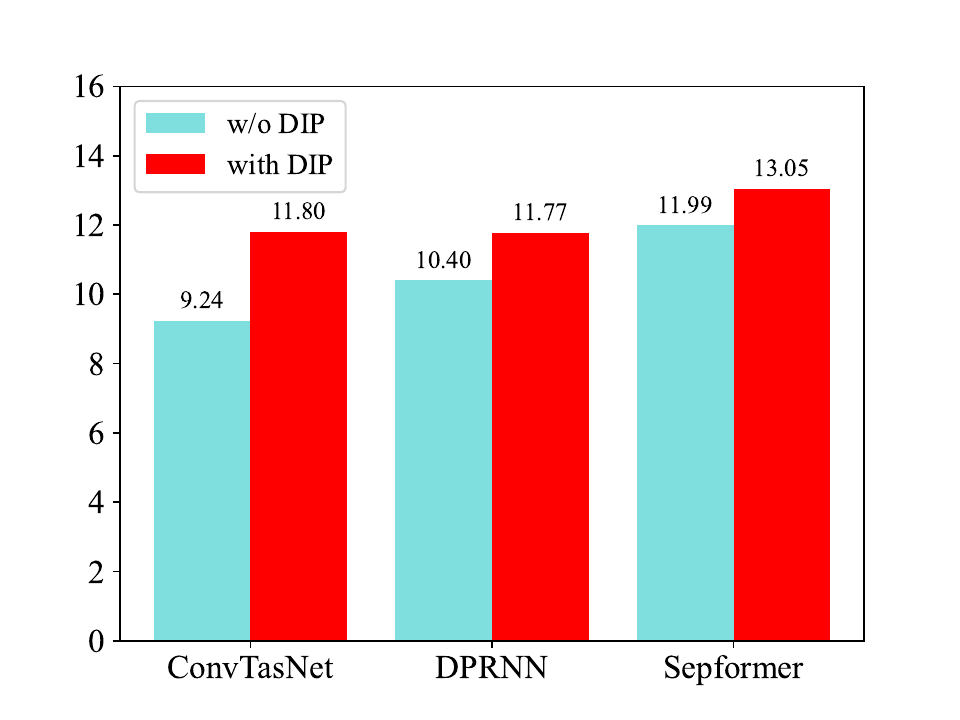}
	\caption{The SI-SDRi(dB) of three different time-domain speech separators to transfer knowledge from LM2Mix to Vox2Mix dataset. The blue bar is the separator without any pretrained frontend and the red bar is the separator with our DIP frontend.} 
	\label{Fig.model_sisdr} 
\end{figure}

\subsection{Integrating with Different Separators}

The experiments in previous sections convincingly demonstrate the effectiveness of our DIP frontend on the ConvTasNet separator. To assess the adaptability of the pretrained frontend, we employ both time-domain and frequency-domain separation models to evaluate the separated speech quality. 

We start to transfer the separation knowledge from the LM2Mix dataset to the Vox2Mix dataset using three different time-domain separators: ConvTasNet, DPRNN, and Sepformer. As depicted in  Fig.~\ref{Fig.model_sisdr}, all three separators with our proposed DIP frontend (red bars) outperform their counterparts without our DIP frontend (blue bars), despite the separator never seeing the labeled data from the Vox2Mix dataset. Remarkably, the improvement provided by our DIP frontend appears to be influenced by the type of separator. It is more obvious for the convolution-based separator (2.47 dB SI-SDRi for the ConvTasNet) and relatively smaller for the attention-based separator (1.15dB SI-SDRi for Sepformer). These results suggest that the convolution-based separation models tend to be overfitting on the train set, and our DIP frontend significantly improves the generalization capacity of these models to other datasets. Additionally, even for state-of-the-art attention-based models, our proposed DIP frontend still can bring sufficient improvement in speech quality.

\begin{table}
	\caption{Experimetn results on SUPERB benchmark to transfer the separation knowledge from the LM2Mix dataset to the Vox2Mix dataset. The Bi-LSTM separator predicts the PSM of input mixture to reconstruct the target reference speech. The \textbf{bold values} are the optimal choices.}
	\label{tab_freq}
	\centering 
	\begin{tabular}{c|c|c|c|c}
		\toprule
		\multirow{2}*{Separator} & \multirow{2}*{Frontend} &  \multicolumn{3}{c}{Test (LM2Mix $\rightarrow$ Vox2Mix)} \\
		~ & ~ & PESQ $\uparrow$ & STOI $\uparrow$ & SI-SDRi(dB) $\uparrow$\\
		\hline
  
		\multirow{8}*{Bi-LSTM} & STFT & 1.38 & 0.76 & 6.13 \\		
		~ & Wav2vec2 & 1.37 & 0.78 & 6.26\\
        ~ & Hubert & 1.37 & 0.78 & 6.20\\
        ~ & Unispeech-sat & 1.40 & 0.80 & 6.81\\
        ~ & WavLM & 1.39 & 0.78 & 6.42\\        
        ~ & Unispeech-sat+ & 1.41 & 0.80 & 7.10\\
        ~ & WavLM+ & 1.43 & 0.81 & 7.34\\        
        ~ & \textbf{DIP} & \textbf{1.43} & \textbf{0.82} & \textbf{7.95}\\
		\bottomrule
	\end{tabular}
\end{table}

Next, we conduct the separation experiment with the frequency-domain separator from SUPERB benchmark, as shown in Table~\ref{tab_freq}. The SUPERB benchmark utilizes the Bi-LSTM separator to predict the PSM for target reference speech. In our domain transfer experiment from the LM2Mix dataset to the Vox2Mix dataset, our proposed DIP frontend obtains the best results in terms of PESQ, STOI, and SI-SDRi metrics. Notably, we find that the improvement of SI-SDRi surpasses that of PESQ and STOI, indicating that our proposed DIP frontend is particularly beneficial for signal accuracy rather than speech coherence.

\subsection{Evaluation on the Real Dataset}

\begin{table}
    \caption{A comparison among different separators with our DIP frontend to transfer the separation knowledge from the LM2Mix dataset to the REAL-M dataset. We implement three types of ASR models to make a comprehensive evaluation with the WER metric. The Sepformer-WHAMR! is the official separation model trained by WHAMR! dataset ~\cite{maciejewski2020whamr}. The \textbf{bold values} are the optimal choices.}
    \centering
    \begin{tabular}{c|c|c|c|c} 
       \toprule
        \multirow{2}*{Separator} & \multirow{2}*{Frontend} & E2E & W2V & WavLM\\
        ~ & ~ & WER $\downarrow$ & WER $\downarrow$ & WER $\downarrow$\\

        \midrule 
        N/A & N/A & 110.20 & 96.47 & 72.63 \\

        \midrule 
        Sepformer-WHAMR! ~\cite{subakan2022real} & N/A & 67.37 & 59.98 & 48.89 \\
        \midrule    
        \multirow{3}*{ConvTasNet} & N/A & 72.19 & 62.22 & 51.08 \\
        ~ & DIP & 53.12 & 45.10 & 36.21 \\
        ~ & \textbf{DIP+} & \textbf{49.74} & \textbf{41.78} & \textbf{32.48} \\

        \midrule    
        \multirow{3}*{DPRNN} & N/A & 66.48 & 58.04 & 45.87 \\
        ~ & DIP & 52.76 & 45.63 & 35.73 \\
        ~ & \textbf{DIP+} & \textbf{41.85} & \textbf{35.73} & \textbf{32.55} \\

        \midrule    
        \multirow{3}*{Sepformer} & N/A & 65.31 & 55.18 & 44.40  \\
        ~ & DIP & 50.61 & 43.63 & 34.17\\
        ~ & \textbf{DIP+} & \textbf{46.98} & \textbf{41.38} & \textbf{31.86}\\
       \bottomrule
    \end{tabular}
    \label{tab_asr}
\end{table}

All previous experiments have consistently demonstrated the capability of our proposed DIP frontend to address the domain gap problem, enabling the downstream model to effectively transfer the separation knowledge from labeled datasets to unlabeled datasets. In this section, we validate our model on the real two-speaker mixtures from the REAL-M dataset(mixture samples except the 'early collection' set), to confirm the benefits of overcoming the domain gap problem for high-quality speech separation in real-world scenarios. Since the real-world data lack target reference speech for direct speech quality evaluation, we employ three pretrained ASR models (E2E transformer, Wav2Vec transformer, and WavLM transformer from the Asteroid and speechbrain toolkit) to evaluate the separated quality indirectly.

Table~\ref{tab_asr} explores two configurations of our proposed DIP frontend: the DIP represents the pretrained frontend in system 13, while the DIP+ denotes the frontend that continually pretrained with 600 mixture waveforms from the 'early collection' set of REAL-M dataset. Notably, leveraging our DIP frontend significantly improves the separated speech quality, consequently reducing the WER of the ASR system. For example, the Sepformer without our DIP frontend achieves 44.40\% WER on the REAL-M dataset. With the DIP frontend, the WER of the Sepformer is reduced by 10.23\% WER (34.17\%). The incorporation of additional real mixtures for frontend pretraining further leads to a 2.29\% WER improvement (31.86\%). Similar trends are observed for the ConvTasNet and the DPRNN separators. These results demonstrate that alleviating the domain gap between real and synthetic samples through our proposed DIP frontend is effective for real-scenario speech separation models to generate high-quality separated speech.

\begin{table}
    \caption{A objective comparison among different separation models by DNSMOS rates to transfer the separation knowledge from the LM2Mix dataset to the REAL-M dataset. The \textbf{bold values} are the optimal choices.}
    \centering
    \begin{tabular}{c|c|c|c} 
       \toprule
        Separator & Frontend & DNSMOS $\uparrow$ & $\widehat{\text{SNR}}$(dB)$\uparrow$\\

        \toprule    
        \multirow{3}*{Sepformer} & N/A & 2.85 & 1.93\\
        ~ & DIP & 2.92 & 2.47\\
        ~ & \textbf{DIP+} & \textbf{2.94} & \textbf{2.49}\\
       \bottomrule
    \end{tabular}
    \label{tab_mos}
\end{table}

We also employ the DNSMOS ratings and neural SNR estimator\footnote{https://huggingface.co/speechbrain/REAL-M-sisnr-estimator} for perceptual quality measurement to evaluate our models.  In contrast to other objective metrics that rely on the availability of the target reference speech, the DNSMOS\footnote{https://github.com/microsoft/DNS-Challenge/tree/master}, trained by the ground truth human
ratings obtained from ITU-T P.808~\cite{reddy2020interspeech}, directly serves as a speech quality estimator. As presented in Table~\ref{tab_mos}, the downstream model benefits from our DIP frontend. In addition, the extra real mixtures in the pretraining corpus further improve 0.02 perceptual quality of separated speech as shown in DIP+ setting when compared with the DIP frontend. Similarly, the neural SNR estimator trained with ground truth SI-SNR scores demonstrates a similar tendency. The Sepformer with our proposed DIP frontend improves by 0.54 dB compared to the baseline system. Utilizing additional real mixtures in the pretraining stage further increases 0.02 dB speech quality. These results clearly
demonstrate that the frontend pretrained by a corresponding corpus is helpful for the downstream model to improve the domain transfer capacity.

\section{Conclusion}
\label{sec:con_ssl}
In this work, we have formulated a conceptually simple and effective DIP frontend that learns domain-invariant contextual cues from speech mixture. By exposing to real-world speech mixture through pre-training, the DIP frontend contributes to the improvement of downstream speech separation tasks consistently. 

\bibliographystyle{IEEEbib}
\bibliography{IEEEabrv,Bibliography}






\vfill

\end{document}